\documentclass{statsoc}
\usepackage[a4paper]{geometry}
\usepackage[utf8]{inputenc}
\usepackage{enumerate} 
\usepackage{amsmath}
\usepackage{graphicx}
\usepackage{diagbox}
\usepackage{multirow}
\usepackage{natbib}
\usepackage{longtable}
\usepackage{color}
\usepackage{comment}
\usepackage{pgf, tikz}
\usepackage{float}
\usetikzlibrary{arrows, automata}

\title[Estimation and Inference for Time-varying Mediation Model]{Estimation and Inference for the Mediation Effect in a Time-varying Mediation Model}

\author{Xizhen Cai}
\address{Department of Mathematics and Statistics, 
Williams College, MA}

\author{Donna L. Coffman}
\address{Department of Epidemiology and Biostatistics, Temple University, PA}

\author{Megan E. Piper}
\address{Department of Medicine, University of Wisconsin, WI}

\author{Runze Li}
\address{Department of Statistics, Pennsylvania State University at University Park, PA}

\author{}

\begin{document}

\begin{abstract}
Traditional mediation analysis typically examines the relations among an intervention, a time-invariant mediator, and a time-invariant outcome variable. Although there may be a direct effect of the intervention on the outcome, there is a need to understand the process by which the intervention affects the outcome (i.e. the indirect effect through the mediator). This
indirect effect is frequently assumed to be time-invariant. With improvements in data collection technology, it is possible to obtain repeated assessments over time resulting in intensive longitudinal data. This calls for an extension of traditional mediation analysis to incorporate time-varying variables as well as time-varying effects. In this paper, we focus on estimation and inference for the time-varying mediation model, which allows mediation effects to vary as a function of time. We propose a two-step approach to estimate the time-varying mediation effect. Moreover, we use a simulation based approach to derive the corresponding point-wise confidence band for the time-varying mediation effect. Simulation studies show that the proposed procedures perform well when comparing the confidence band and the true underlying model. We further apply the proposed model and the statistical inference procedure to real-world data collected from a smoking cessation study. 
\end{abstract}


\maketitle

\section{Introduction}
Developments in mobile and wearable device technology have enabled the collection of intensive longitudinal data \citep{schafer2006models}, such as ecological momentary assessment (EMA) \citep{shiffman2008, shiffman2009}, in which data on variables such as craving, withdrawal symptoms, or stress, are measured in real-time, real-world contexts. EMA is particularly useful in health behavior change studies, for example, smoking cessation studies. Often, the variables that are collected during EMA are variables that are targets of a behavior change intervention and are also thought to affect the health outcomes of interest. In other words, they are mediators, variables that lie on the pathway between the intervention and the outcome.


As the collection of data using EMA has grown, so have methods for analyzing and making the most of the temporal density of measurements, such as the mixed-effects location scale model \citep{nordgren2020extending} and the time-varying effect model \citep{tan2012}. However, there have been very few proposed methods for assessing mediation using this type of intensively measured data. This paper aims to propose an approach to mediation in which both the mediator and outcome are collected using EMA, and therefore can be examined as they change over time. Specifically, we first proposed a two-step approach to estimate the time-varying mediation effect. 
To make statistical inference on the time-varying mediation effect, we develop a simulation-based approach to derive the corresponding point-wise 
confidence band for the time-varying mediation effect.

The rest of this paper is organized as follows. In section \ref{section:varying}, we present relevant background material on varying-coefficient models and the proposed model for time-varying mediation, including estimation and bootstrap inference. In section \ref{section:simulations}, we present simulation studies to examine the performance of the bootstrap confidence intervals. In section \ref{section:application}, we apply the proposed methods to data from a smoking cessation intervention study. In section \ref{section:discussion}, we discuss limitations, future directions, and conclusions.

\section{Varying-Coefficient Models and Proposed Extension to Mediation} \label{section:varying}
Time-varying coefficient models \citep{hoover1998nonparametric} have been used to model time-varying effects of an independent variable on a dependent variable \citep{tan2012, dziak2012mixture, dziak2014ordinal}. These are essentially varying-coefficient models \citep{hastie1993varycoef} applied to intensive longitudinal data. For each individual, $i$, the independent variable and the outcome variable are measured at multiple time points $\{t_{ij}, j=1,2,\ldots, T_i\}$. The data collected are 
$$\{t_{ij}, X_i(t_{ij}), Y_i(t_{ij})\}, \quad\mbox{for} \quad i=1,2,\ldots,n,\quad  j=1,2,\ldots,T_i$$
and the model can be written as 
$$Y_{i}(t_{ij})={\beta}_{0}(t_{ij})+X_i(t_{ij}){\beta}_{1}(t_{ij})+ \epsilon_i(t_{ij}),$$
where ${\beta}_{0}(t)$ and $\beta_1(t)$ are smoothing functions of time, and therefore are called the time-varying coefficient functions, and $\epsilon(t)$ is a zero-mean stochastic process with covariance function, $\gamma(s,t)$. Not only are the effects (i.e., coefficients) of the predictor variables time-varying, but the values of the variables themselves can also change over time. Different types of estimation procedures have been well summarized \citep{fan1999vary}. There are essentially two estimation approaches for time-varying effect models: splines and local smoothing methods. In this paper, we focus on 
local smoothing methods, which locally approximate coefficient functions by linear or polynomial functions \citep{fan1996local}.

\cite{fan2000two} proposed a powerful two-step procedure that uses kernel methods to estimate the time-varying coefficients and their corresponding standard errors. Both simulations and real data applications showed the efficiency of their method over other previous proposals. Later,  \cite{csenturk2008generalized} extended the model and its estimation to the case where the outcome variable depends not only on the current but also past values of the independent variables, 
$$Y_{ij}(t_{ij})=\beta_0(t_{ij})+\sum_{r=1}^{p}\beta_{r}(t_{ij}){X}_{i,j-q-(r-1)}(t_{i,j-q-(r-1)})+\epsilon_{ij}(t_{ij}),$$ where $p$ is the number of past time points that are assumed to affect the current response and $q$ incorporates the possible time lag. The estimation is done through a variant of the two-step procedure and has been shown to perform well. This two-step procedure \citep{fan2000two} provides an important foundation for our proposed estimation procedure for time-varying mediation effects which combines the traditional linear mediation model estimation procedure and local polynomial smoothing.

Although time-varying coefficient models are relatively common for examining the time-varying effect of an independent variable on a dependent variable, relatively little work has examined time-varying effects for mediation.  \cite{lindquist2012functional} first introduced functional (or time-varying) mediation effects in which the independent and dependent variables were measured at a single point in time but the mediator was measured intensively over time using fMRI. More recently, \cite{vanderweele2017mediation} proposed a mediation g-formula, which allows time-varying treatments, time-varying mediators, and an end-of-study point outcome. They mention the possibility of time-varying effects, but their study does not directly address them. In our application to a smoking cessation study, the mediator and outcome are both measured repeatedly over time (i.e., time-varying mediator and time-varying outcome) and the independent variable is random assignment to the intervention (not a time-varying treatment). Thus, neither of these previous models apply directly to our smoking cessation study.

\begin{figure}[ht!]
\begin{center}
\begin{tikzpicture}[
            > = stealth, 
            shorten > = 1pt, 
            auto,
            node distance =3cm, 
            semithick 
        ]

        \tikzstyle{every state}=[
            draw = black,
            thick,
            fill = white,
            minimum size = 7mm
        ]

        \node[draw] (X) {Intervention ($X$)};
        \node[draw] (M) [above right of=X] {Mediator ($M$)};
        \node[draw] (Y) [below right of=M] {Outcome ($Y$)};

        \path[->] (X) edge node {$\alpha$} (M);
        \path[->] (X) edge node {$\gamma$} (Y);
        \path[->] (M) edge node {$\beta$} (Y);

    \end{tikzpicture}
\end{center}
\caption{The traditional mediation model with time-invariant effects}\label{fig:TraditionalMediationModel}
\end{figure}
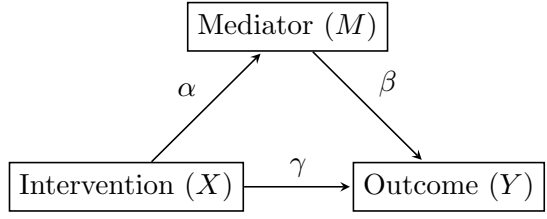

Traditional methods of assessing mediation, shown in Figure \ref{fig:TraditionalMediationModel}, generally specify the direct effect (i.e., that does not go through the mediator) as $\gamma$, and the indirect or mediated effect as the product of paths $\alpha$ (effect of intervention on mediator) and $\beta$ (effect of mediator on outcome). The standard error of this product term, $\alpha\beta$, is obtained either asymptotically \citep{sobel1982asymptotic} or via bootstrap procedures in order to test the statistical significance of the mediated effect. Several simulation studies have shown bootstrap standard errors to be superior, especially in smaller samples because $\alpha\beta$ may not be normally distributed \citep{mackinnon2002compare, shrout2002mediation}.

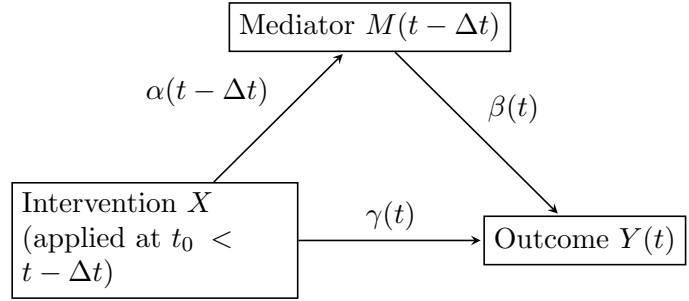
\begin{figure}[ht!]
\begin{center}
\begin{tikzpicture}[
            > = stealth, 
            shorten > = 1pt, 
            auto,
            node distance =4cm, 
            semithick 
        ]

        \tikzstyle{every state}=[
            draw = black,
            thick,
            fill = white,
            minimum size = 7mm
        ]

        \node[draw] (X) [text width=3.5cm] {Intervention $X$ \\ (applied at $t_0 < t -\Delta t$)};
        \node[draw] (M) [above right of=X] {Mediator $M(t-\Delta t)$};
        \node[draw] (Y) [below right of=M] {Outcome $Y(t)$};

        \path[->] (X) edge node {$\alpha (t- \Delta t)$} (M);
        \path[->] (X) edge node {$\gamma(t)$} (Y);
        \path[->] (M) edge node {$\beta(t)$} (Y);

    \end{tikzpicture}
\end{center}
\caption{The proposed time-varying mediation model}\label{fig:TimeVMediationModel}
\end{figure}

As in traditional mediation analysis with time-invariant effects, the time-varying mediation effect is the product of two effects, but in this case, both effects are time-varying. That is, the two effects are no longer single numbers such as $\alpha$ and $\beta$; rather, they are functions of time, and the product term is also a function of time. Figure \ref{fig:TraditionalMediationModel} is extended in Figure \ref{fig:TimeVMediationModel} to include time-varying effects.


In this paper, we propose to estimate the time-varying mediation model by extending the two-step approach \citep{fan2000two}, followed by bootstrapping to obtain confidence intervals for the indirect effect (i.e., the effect of the intervention on the outcome through the mediator). Mediation is inherently about causal pathways - the intervention changes the mediator, which in turn has an effect on the outcome. To infer causality, we will need to assume that there are no unmeasured confounders, additivity (no interactions or non-linearities), and no time-varying confounders. In addition, we assume temporal ordering such that the intervention occurs before the mediator which occurs before the outcome. These are the standard assumptions needed for a linear structural equation model to estimate a ``causal" effect. These are strong assumptions, to which we return when discussing future directions.

\subsection{The proposed model}
Extending the mediation framework in Figure \ref{fig:TraditionalMediationModel} to take advantage of the temporal density of intensive longitudinal data allows us to estimate time-varying effects as shown in the dynamic mediation diagram in Figure \ref{fig:TimeVMediationModel}. In this model, we consider the measurement timing of the variables consistent with modeling mediation as a process that unfolds over time (i.e., intervention must precede change on the mediator, and mediator must precede change on the outcome). The intervention or independent variable, denoted $X$, is time-invariant, and assigned at time $t_0$. Across time, the effect of $X$ on the value of the mediator $M$ at any time $t>t_0$ (i.e., $M(t)$) is denoted by $\alpha(t)$. The value of the outcome variable $Y$ at time $t$ (i.e., $Y(t)$) is affected by the value of the mediator at a small window before time $t$, i.e., $M(t-\Delta t)$. Here $\Delta t$ is a small constant which represents the time-lag of the effect of the mediator on the outcome. More discussion on the choice of $\Delta t$ will be presented shortly.

The diagram in Figure \ref{fig:TimeVMediationModel} leads to the following time-varying mediation model: 
\begin{eqnarray*}
M(t) &= &  \alpha_0 (t)+I (t \geq t_0) \alpha (t)   X+ \epsilon^{M}(t),\\
Y(t)  &=&  \beta_0 (t)+I (t \geq t_0) \left\{\gamma(t)  X + \beta(t) M(t-\Delta t) \right\}+ \epsilon^{Y}(t),
\end{eqnarray*}
where $ \epsilon^{M} (t)$ and $\epsilon^{Y} (t)$ are both zero-mean stochastic processes. The time-varying mediation effect of interest is $\alpha(t-\Delta t)\beta(t)$. Suppose there are repeated measurements of $N$ subjects at multiple time points $\{t_{ij}\}$, then the observed data are 
$$\{X_i, (t_{ij}, M_i(t_{ij}), Y_i(t_{ij}))\}, \quad i=1,2,\ldots, N, \quad  j=1,2,\ldots,T_i,$$
and the model is
\begin{eqnarray*}
M_{i}(t_{ij}) &= &  \alpha_0 (t_{ij})+I (t_{ij} \geq t_0) \alpha (t_{ij})   X_i+ \epsilon^{M}_i (t_{ij})\\
Y_{i}(t_{ij})  &=& \beta_0 (t_{ij})+I (t_{ij} \geq t_0) \left\{\gamma (t_{ij})  X_i + \beta(t_{ij}) M_{i}(t_{ij} -\Delta t) \right\}+ \epsilon^{Y}_i (t_{ij}).
\end{eqnarray*}
Note that all effects of the intervention, $X_i$, are controlled by a post-intervention indicator, $I (t_{ij} \geq t_0)$, because the intervention is assigned at time $t_0$. Since the mediation (i.e., indirect) effect is of primary interest, we focus on the time points after the intervention is assigned, thus, the indicator term can be dropped and the final model is, 
\begin{eqnarray}
M_{i}(t_{ij}) &= &  \alpha_0 (t_{ij})+ \alpha (t_{ij})   X_i+ \epsilon^{M}_i (t_{ij}) \label{eqn:finalmodel1} \\
Y_{i}(t_{ij})  &=& \beta_0 (t_{ij})+ \gamma (t_{ij})  X_i + \beta (t_{ij}) M_{i}(t_{ij}-\Delta t)+  \epsilon^{Y}_i (t_{ij}). \label{eqn:finalmodel2} 
\end{eqnarray}

To use all information in the collected data effectively, the model in equation (\ref{eqn:finalmodel2}) indicates that the window size $\Delta t$ should be smaller than the distance between the current time point $t_{ij}$ and the previous time point $t_{i,j-1}$. If the observation times are not equally spaced, the window size can be chosen to be a value that is smaller than the smallest distance. In our simulation and applied analysis, we assume that the value of the mediator at time $t_{ij}-\Delta t$ can be substituted by its value at the previous time point $M_{i}(t_{i,j-1})$. Additionally, models (\ref{eqn:finalmodel1})
and (\ref{eqn:finalmodel2}) assume that there are only two intervention groups (e.g., treatment versus control) such that $X_i$ is a binary indicator of the treatment condition, and the time-varying effect $\alpha(\cdot)$ is the effect of the treatment as compared to the control group. For more than two intervention groups, the proposed model can be easily extended by adding more indicator variables (see the smoking cessation study in Section \ref{section:application} as an example). Without loss of generality, we present the following proposed estimation procedure and bootstrap inference for the models in equations (\ref{eqn:finalmodel1}) and (\ref{eqn:finalmodel2}). 

\subsection{Estimation of the time-varying mediation effect}

We could estimate the time-varying effects in model equations (\ref{eqn:finalmodel1})  and  (\ref{eqn:finalmodel2})  separately using the two-step estimation procedure \citep{fan2000two}. Here we propose a variant of that approach to estimate them simultaneously. 

Let $\{t_1, t_2, \ldots, t_T\}$ be the distinct time points when the data are measured.  For any fixed time point $t_j\in\{t_2, \ldots, t_T\}$, we observe complete data from $N_j$ subjects ($N_j$ does not necessarily equal $N$). Then for any individual $i$ at this fixed time point $t_j$, the observed data are 
$$
(X_i, M_{ij}, Y_{ij}), \quad i=1,2,\ldots, N_j,
$$
\noindent where $M_{ij}=M_i(t_{ij})$ and $Y_{ij}=Y_i(t_{ij})$. Similar to the first step of the two-step procedure \citep{fan2000two}, at any fixed time $t_j$, model equations (\ref{eqn:finalmodel1})  and  (\ref{eqn:finalmodel2})  become the traditional linear mediation model. So we can estimate the value of the varying coefficient functions $\alpha(t_j), \beta(t_j)$, and $\gamma(t_j)$, which are treated as three parameters rather than three functions, by the least squares method, namely, by solving the following two optimization problems,
$$\min_{\alpha} \sum_{i=1}^{N_j} (M_{ij}-\alpha(t_j) X_i)^2 \mbox{ and }\min_{\beta, \gamma} \sum_{i=1}^{N_j} (Y_{ij}-\gamma(t_j) X_i - \beta(t_j) M_{i,j-1})^2.$$
Suppose the outcome and independent variables are mean centered or standardized at the fixed time point so that we can safely drop the intercept terms. To derive a joint distribution of the estimated coefficients, we propose to combine the two least squares problems together to create a new least squares problem given as
\begin{eqnarray}
&& \min_{\alpha, \beta, \gamma} \left\{\sum_{i=1}^{N_j} (M_{ij}-\alpha (t_j) X_i)^2 + \sum_{i=1}^{N_j} (Y_{ij}-\gamma (t_j) X_i - \beta (t_j) M_{i,j-1})^2\right\} \label{eqn:leastsquares} \\ 
&\Leftrightarrow& \min_{\boldsymbol{\delta}}\sum_{i=1}^{2N_j} (Y^{*}_{ij}-\boldsymbol{\delta}^{\top}(t_j)\boldsymbol{X}^{*}_{ij})^2 \nonumber
\end{eqnarray}
where $\boldsymbol{\delta}(t_j)=(\alpha(t_j), \gamma(t_j), \beta(t_j))^{\top}$, and $Y^{*}_{ij}$, $X^{*} _{ij}$ in matrix forms are, 
\begin{eqnarray*}
\mathbf{Y}^{*} _j = \left(\begin{array}{c}M_{1j}\\M_{2j} \\ \vdots\\M_{N_j,j}\\ Y_{1j} \\ Y_{2j} \\ \vdots \\ Y_ {N_j,j}\end{array}\right)_{2N_j\times 1},
\mathbf{X}^{*} _j = \left(\begin{array}{ccc}X_{1} & 0 & 0 \\X_{2} & 0 & 0  \\ \vdots & \vdots& \vdots \\ X_{N_j} & 0 & 0 \\ 0 & X_{1} & M_{1,j-1} \\ 0 & X_{2} & M_{2,j-1} \\ \vdots & \vdots& \vdots  \\ 0 & X_{N_j} & M_{N_j,j-1}  \end{array}\right)_{2N_j\times 3}
\end{eqnarray*}
Denote the solution to the least squares problem in (\ref{eqn:leastsquares}) as $\boldsymbol{d} (t_j) = (a(t_j), c(t_j),  b(t_j))^{\top}$,  i.e. $\boldsymbol{d}(t_j)$ is an estimate of $\boldsymbol{\delta}(t_j)$ and is a $3\times (T-1)$ dimensional vector, which includes values of the estimated time-varying coefficient functions at all time points, 
\begin{eqnarray*}
\boldsymbol{d}=\left(a(t_2),c(t_2),b(t_2), a(t_3),c(t_3),b(t_3), \cdots, a(t_T),c(t_T),b(t_T)\right)^{\top}
\end{eqnarray*}

Similar to the second step of the two-step procedure \citep{fan2000two}, the smoothed coefficient functions $\hat{\alpha}$ and $\hat{\beta}$ in model equations (\ref{eqn:finalmodel1}) and  (\ref{eqn:finalmodel2}) are further calculated by local polynomial regression using $(a(t_j), b(t_j))$ as, 
\begin{eqnarray}
\hat{\alpha} (t- \Delta t) &=& \sum _{l=2}^T w(t_l, t- \Delta t) a (t_l)\\
\hat{\beta} (t) &=& \sum _{l=2}^T w(t_l, t) b (t_l)
\end{eqnarray}
where $w(t_j, t)$ can be weights from any linear smoothing techniques. Here, we use local polynomial weights. Then the desired mediation effect is 
\begin{eqnarray}
\hat{\alpha} (t- \Delta t)\hat{\beta} (t)  = \left\{\sum _{l=2}^T w(t_l, t- \Delta t) a (t_l) \right\}\left\{\sum _{l=2}^T w(t_l, t) b (t_l)\right\}, \label{eqn:mediationeffect}
\end{eqnarray}
and each part can be rewritten as linear combinations of $\boldsymbol{d}$: 
\begin{eqnarray}
\hat{\alpha} (t- \Delta t)\hat{\beta} (t)  = (\mathbf{w}_a^T\mathbf{d})(\mathbf{w}_b^T\mathbf{d}),
\label{eqn:linearcombination}
\end{eqnarray}
where

$$\mathbf{w_a}=
\left(\begin{array}{c}
w(t_2, t- \Delta t)\\
0\\
0\\
w(t_3, t- \Delta t)\\
0\\
0\\
\vdots\\
w(t_T, t- \Delta t)\\
0\\
0\\
\end{array}\right), \mbox{and } \mathbf{w_b}=\left(\begin{array}{c}
0\\
0\\
w(t_2, t)\\
0\\
0\\
w(t_3, t)\\
\vdots\\
0\\
0\\
w(t_T, t)\\
\end{array}\right).$$

\subsection{Estimating point-wise confidence interval of mediation effect through bootstrap}


To identify a statistically significant mediation effect in the time-varying setting, we consider the following hypothesis:

\begin{eqnarray*}
&&H_0:{\alpha} (t- \Delta t){\beta} (t) = 0\mbox{ for any fixed }t\\
&vs.& H_A: \mbox{the mediation effect is not zero}.
\end{eqnarray*}
Since the distribution of the mediation effect is not necessarily normal, we use a bootstrap approach to construct confidence intervals. Specifically, the lower and upper bounds of the $1-\alpha$\% confidence interval are taken to be the corresponding lower ($\alpha/2$) and upper ($1-\alpha/2$) percentiles of the distribution of the estimated mediation effect from a large number of bootstrapped samples \citep{efron1994bootstrap}.

For any fixed time $t$, the above bootstrap percentile method creates a point-wise confidence interval for the mediation effect at that $t$. Connecting all confidence intervals yields a confidence band, but this is different from a simultaneous confidence band throughout the entire time interval, since  the nominal confidence level is only satisfied at each fixed time point $t$. We return to this point in Section \ref{section:discussion}. The estimation procedure and bootstrapped confidence intervals are implemented in an R package, \emph{tvmediation}, that is available on GitHub.



\section{Simulation Studies}\label{section:simulations}
To examine the performance of the proposed point-wise confidence interval, we consider the following two simulation models,
\begin{enumerate}[i.]
\item 
$\alpha_1(t)=10+12t^3$, $\gamma(t)=-20-18t$,
$\beta(t)=50+150t^2$, $\gamma(s,t)=15\exp(-0.3|s-t|)$

\item 
$\alpha(t)=15+8.7\sin(0.5\pi t)$,
$\gamma(t)=4-17(t-1/2)^2$,
$\beta(t)=1+2t^2+11.3(1-t)^3$,
 $\gamma(s,t)=15\exp(-0.3|s-t|)$
\end{enumerate}

\noindent The first model includes polynomial functions of different orders, and the second model incorporates a $\sin$ function to increase the complexity of the mediation effect. The two models are similar to those in \cite{fan2000two} and \cite{csenturk2008generalized}

Without loss of generality, observation times are generated as $50$ equally spaced time points between $0$ and $1$. And the time lag $\Delta t$ is chosen to be half of the length between any two consecutive time points. To generate the simulated data, we first randomly assign intervention and control group, each with probability of $0.5$. The error term is generated from multivariate normal distribution with mean zero and covariance $\gamma(s, t)$. The value of the mediator and the outcome variables are generated according to equations (\ref{eqn:finalmodel1}) and (\ref{eqn:finalmodel2}). In the second step of the estimation procedure, local linear regression is used and the bandwidth is chosen by the rule of thumb formula in  section 4.2 of \cite{fan1996local}. 

We consider three simulation settings with $N=100, 200$, and $500$, separately. To verify the nominal level for $95\%$ confidence intervals, we calculated the coverage rate of the proposed point-wise confidence interval at $t=0.2, 0.4, 0.6$, and $0.8$, separately. Table~\ref{tab1} summarizes the results based on $500$ simulation replications.

\begin{table}
    \caption{\label{tab1}Coverage rate for $95\%$ confidence intervals}
\centering
\fbox{%
\begin{tabular}{cc | cccc }
\hline
               & Sample  &  \multicolumn{4}{c}{Coverage}  \\ 
               & Size &  t=0.2 & t=0.4 & t=0.6 & t=0.8 \\
               \hline
        & 100 &  0.954 & 0.956 & 0.944 & 0.950 \\ 
Model i & 200 & 0.948 & 0.956 & 0.958  & 0.954 \\ 
        & 500 & 0.956  & 0.948 & 0.954 & 0.956\\
\hline
        & 100 &  0.946 & 0.952 & 0.946 & 0.948 \\
Model ii & 200 &  0.944 & 0.934&  0.930 & 0.934\\
        & 500 & 0.948 & 0.950 & 0.960 & 0.938 \\
\hline 
\end{tabular}}
\end{table}

\begin{figure}[ht!]
\centering
\includegraphics[scale=0.45]{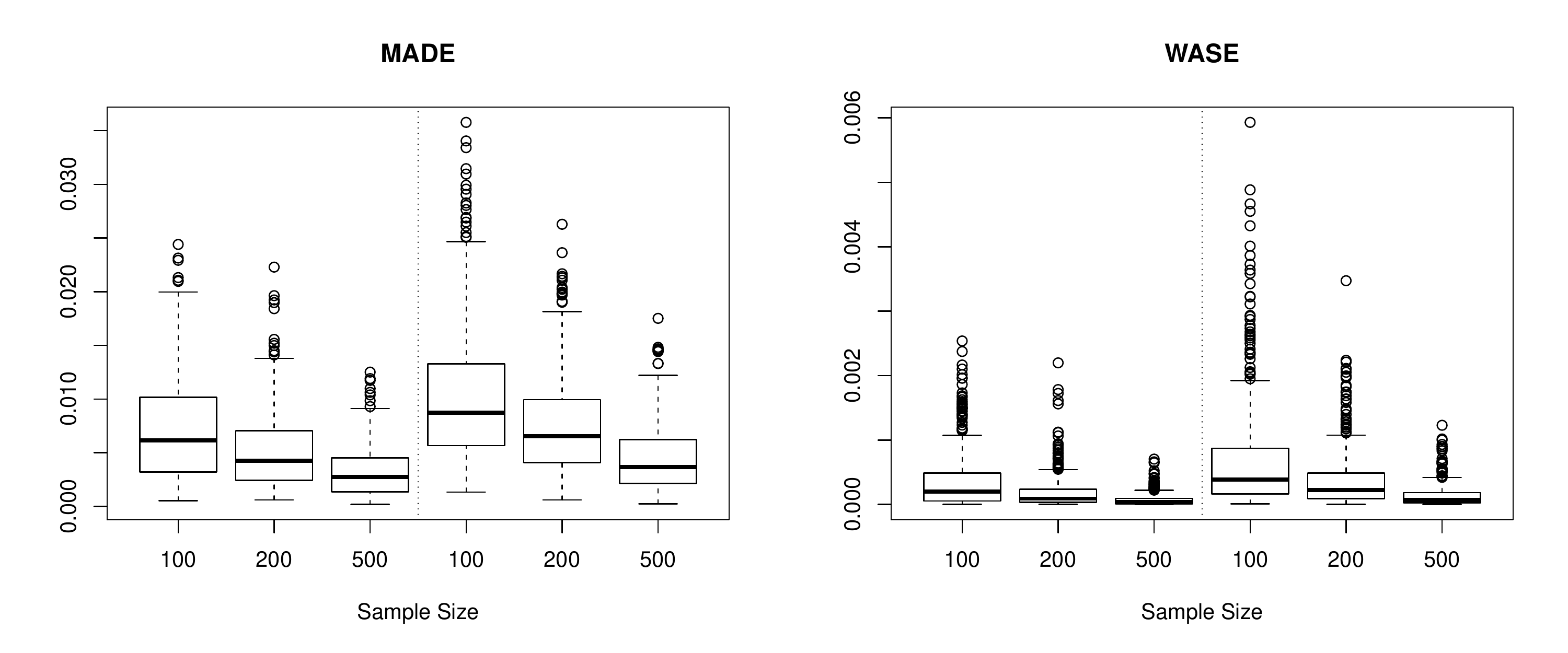}
\caption{MADE and WASE for two models with different sample sizes. In each graph, the left three boxplots are for model i and the right ones are for model ii.}\label{fig:WASE}
\end{figure}

Except for a few settings, the coverage rates are all near a $95\%$ confidence level at these points. We also evaluated the performance of the estimated time-varying mediation effect by the mean absolute deviation error (MADE) and weighted average squared error (WASE) \citep{fan2000two, csenturk2008generalized}, defined as follows,

\begin{equation*}
MADE =  (4T)^{-1} \sum_{j=1}^{T} \frac{|\eta(t_j)-\hat{\eta}(t_j)|}{\mbox{range}(\eta)}, \quad
WASE = (4T)^{-1} \sum_{j=1}^{T} \frac{\{\eta(t_j)-\hat{\eta}(t_j)\}^2}{\mbox{range}^2(\eta)}
\end{equation*}

\noindent where $\eta(t)={\alpha} (t- \Delta t){\beta (t)}$ is the time-varying mediation effect. 
Figure \ref{fig:WASE} presents boxplots of these measurements for the two models. Not surprisingly, both of them show similar patterns with different sample sizes and models.  Specifically, both MADE and WASE decrease as sample size increases for a particular model, and the error for model ii is slightly higher than that of model i.

\begin{figure}[ht]
\centering
\includegraphics[scale=0.35]{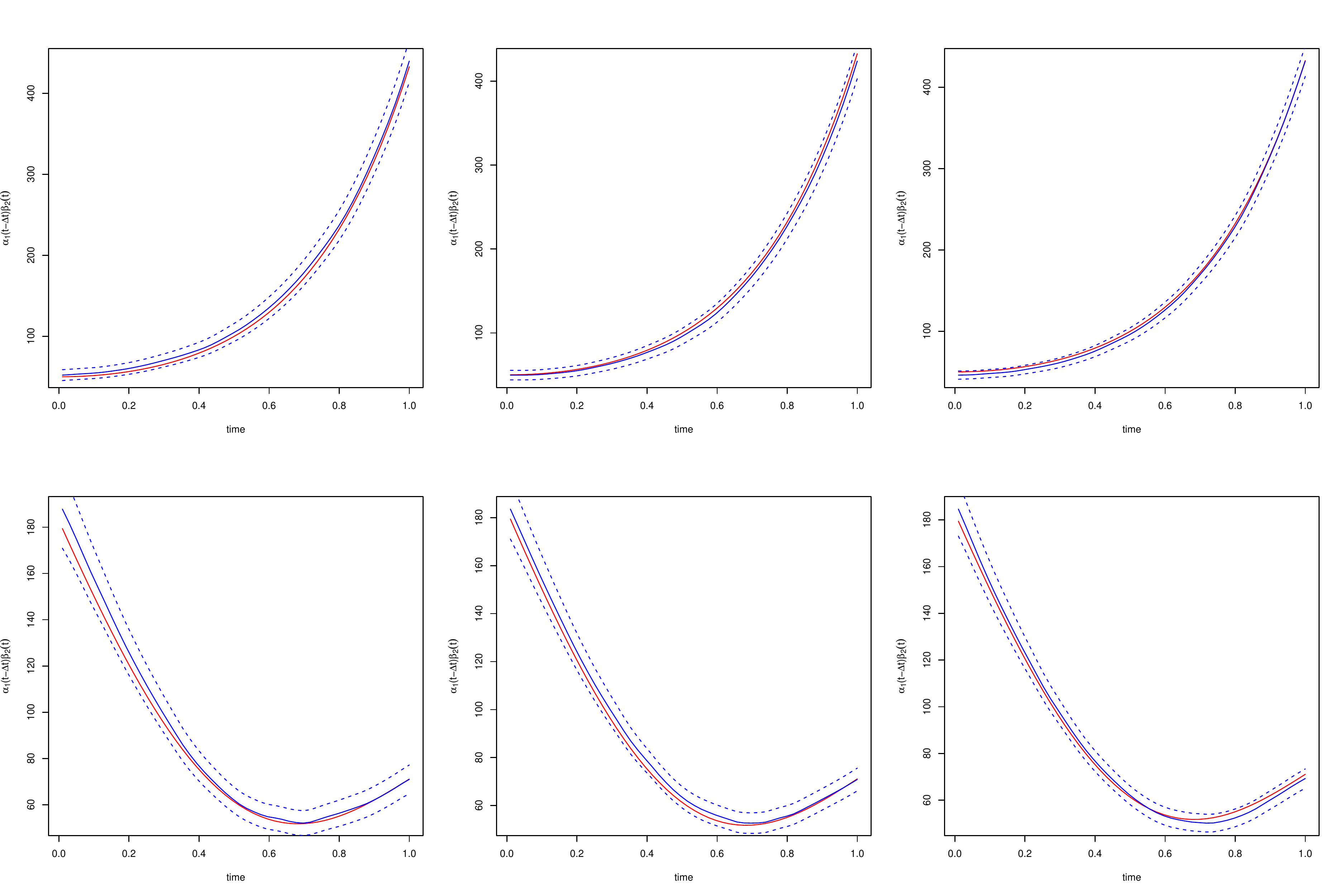}
\caption{Point-wise confidence band of the time-varying mediation effect for the two simulation models. Plots in top row are for simulation model i, where the sample sizes are $100, 200$, and $500$ from left to right. Plots in the bottom row are similar, but for simulation model ii.}\label{fig:simulation}
\end{figure}

To present a typical fit of the proposed procedure, we selected the simulation sample with MADE closest to the median value among all $500$ replications.  The estimated time-varying mediation effect and the  corresponding confidence intervals are plotted in Figure \ref{fig:simulation}, as compared to the true effect. 

The red solid line is the true time-varying mediation effect, and the blue solid line is the estimated effect. For both models, the estimated effect is close to the true underlying effect. The blue dashed lines are the limits of the point-wise confidence band estimated by the proposed method. For both models, the width of the confidence band is not constant throughout the whole time range, but at each time, the true effect is fully contained in the confidence band. As the sample size increases, the confidence band becomes narrower.

\section{Application: The Wisconsin Smokers' Health Study 2} \label{section:application}

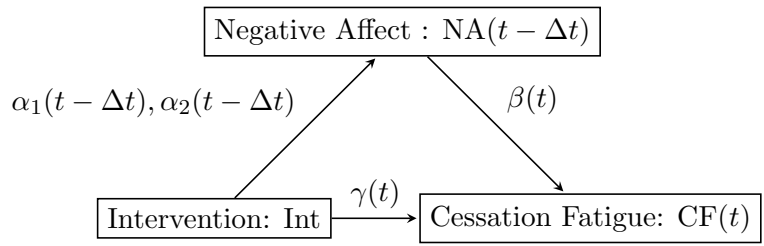
\begin{figure}
\centering
\makebox{
\begin{tikzpicture}[
            > = stealth, 
            shorten > = 1pt, 
            auto,
            node distance =3.5cm, 
            semithick 
        ]
        \tikzstyle{every state}=[
            draw = black,
            thick,
            fill = white,
            minimum size = 4mm
        ]

        \node[draw] (Intervention) {Intervention: Int};
        \node[draw] (Negative Affect) [above right of=Intervention] {Negative Affect : NA$(t-\Delta t)$};
        \node[draw] (Cessation Fatigue) [below right of=Negative Affect] {Cessation Fatigue: CF$(t)$};
        \path[->] (Intervention) edge node {$\alpha_1(t-\Delta t), \alpha_2(t-\Delta t)$} (Negative Affect);
        \path[->] (Intervention) edge node {$\gamma(t)$} (Cessation Fatigue);
        \path[->] (Negative Affect) edge node {$\beta(t)$} (Cessation Fatigue);
\end{tikzpicture}}
\caption{\label{fig:RealdataModel2}The time-varying mediation effect model for the smoking cessation study}
\end{figure}

We applied the proposed method to conduct an empirical analysis of data collected from a smoking cessation study, the Wisconsin Smoker's Health Study 2 \citep{baker2016}, which used EMA to assess negative affect and cessation fatigue during a smoking cessation trial. The study was a randomized comparative efficacy trial \citep{baker2016} directly comparing the two most effective smoking cessation therapies (varenicline vs. combination nicotine replacement therapy - nicotine patch + nicotine mini-lozenge) with one another and with an active comparator treatment (nicotine patch only). In total, $1086$ smokers recruited from Madison and Milwaukee, WI were randomly assigned to one of the three $12$-week pharmacotherapies. Participants completed one morning EMA prompt, and one evening prompt every day for one week prior to the quit day and for two weeks after the quit day and then every other day for the remaining two weeks of the EMA period (i.e., total of one week pre-quit and four weeks post-quit). Thus, there are $14$ EMAs prior to the quit day and $42$ after the quit day.  The goal of our empirical analysis is to examine whether the intervention has an effect on cessation fatigue that is mediated by negative affect (see Figure \ref{fig:RealdataModel2} for the mediation diagram).  

Cessation fatigue, defined as tiredness of trying to quit smoking \citep{piasecki2002have}, and negative affect, measured by asking participants if they were in a negative mood in the last 15 minutes, were both measured on 7-point Likert scales. Previous studies have found that negative affect and cessation fatigue are positively related and related to cessation failure \citep{liu2013understanding}.

\begin{figure}[ht]
\centering 
\includegraphics[scale=0.6]{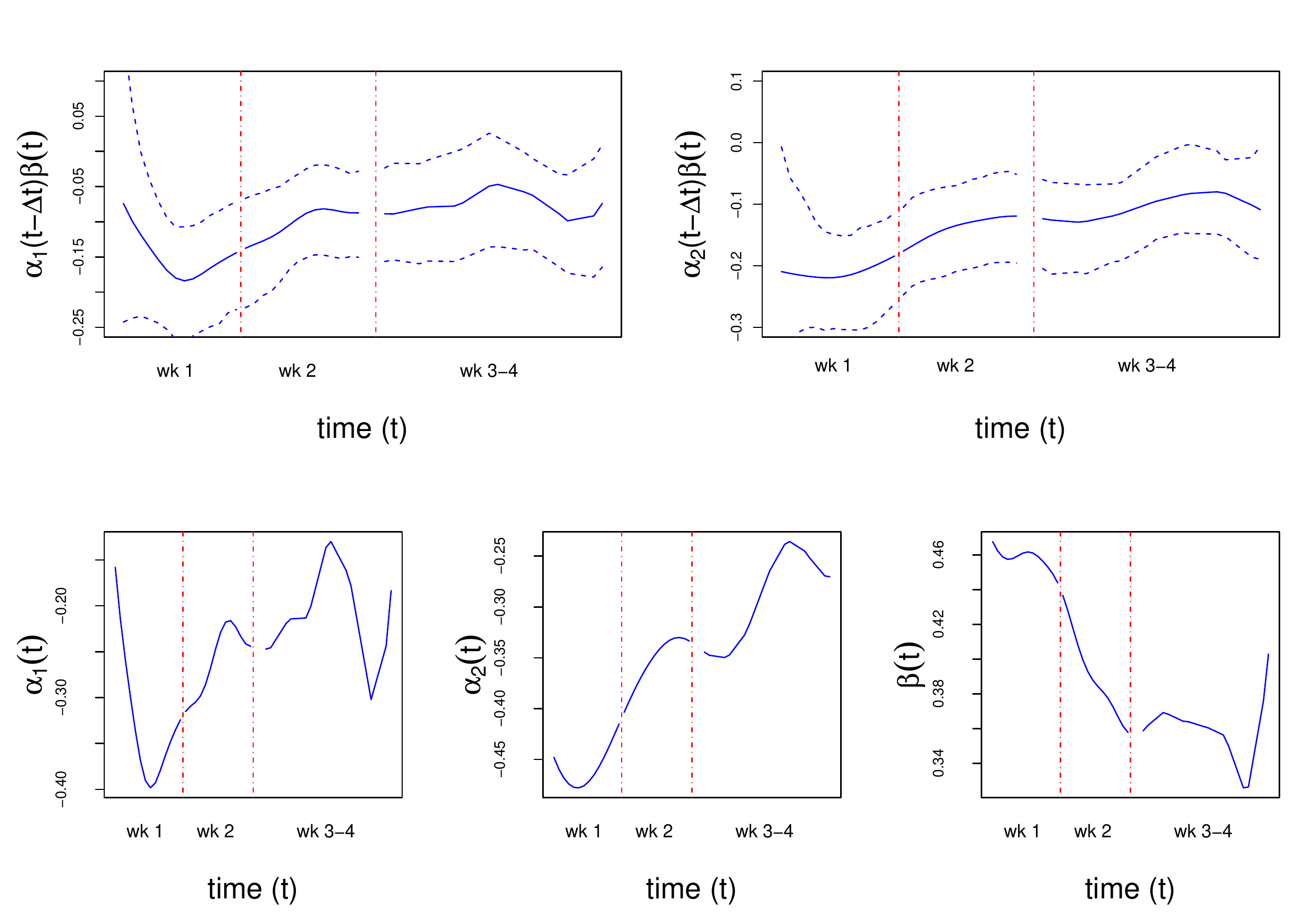} \caption{Time-varying mediation effects and individual effects. The top two plots display the mediation effects with the corresponding point-wise confidence intervals. Left panel is for the treatment varenicline, and the right panel is for the treatment combination nicotine replacement therapy, both as compared to the treatment of nicotine patch alone. The red vertical lines are separation of weeks. The three plots on the bottom row display the individual time-varying effects in the mediation model \ref{eqn:smoking1} and \ref{eqn:smoking2}.}\label{fig:RealdataAnalysis2}
\end{figure}

We use data from the $42$ EMAs after the quit day. Both the time-varying outcome, cessation fatigue, and time-varying mediator, negative affect, are assessed at each EMA prompt. Unlike in the simulation studies, and as is common in most empirical studies, especially with wearable and mobile devices, there are intermittent missing values in the data. Excluding individuals with no data at all, we have 1047 individuals in total, and the observed data are
$$\{\mbox{Varen}_i, \mbox{cNRT}_i,(t_{ij}, \mbox{NA}_{ij}, \mbox{CF}_{ij})\},\quad i=1,2,\ldots,1047, \quad j=1,2,\ldots,42.$$
There are two indicator variables for the intervention: Varen$_i$ indicates assignment to the varenicline group and cNRT$_i$ indicates assignment to the combination nicotine replacement therapy group. The nicotine patch only condition is the reference group as it is considered the standard of care. Additionally, the observation times are not equally spaced (i.e. everyday for the first two weeks and every other day for the remaining two weeks). The previous proposed model can be modified to incorporate the additional intervention condition as follows:
{\small
\begin{eqnarray}
NA_{i}(t_{ij}) &= &  \alpha_0 (t_{ij})+ \alpha_{1}(t_{ij})   Varen_i+ \alpha_{2}(t_{ij})   cNRT_i + \epsilon^{NA }_i (t_{ij}) \label{eqn:smoking1} \\
CF_{i}(t_{ij})  &=& \beta_0 (t_{ij})+ \gamma_{1} (t_{ij}) Varen_i + \gamma_{2}(t_{ij}) cNRT_i + \beta(t_{ij}) NA_{i}(t_{ij}-\Delta t)+  \epsilon^{CF}_i (t_{ij}) \label{eqn:smoking2}.
\end{eqnarray}}
Using the proposed method, the estimated mediation effects, $\alpha_1(t-\Delta t)\beta(t)$ and $\alpha_2(t-\Delta t)\beta(t)$, with the corresponding confidence bands are presented in Figure \ref{fig:RealdataAnalysis2}. 

As compared to the nicotine patch only, the effect of varenicline on cessation fatigue that is mediated by negative affect becomes more negative shortly after quitting. More specifically, the {\it magnitude} of this negative mediation effect increases quickly in the first week after quit day, has a slower decrease in the second week, becomes stable in week 3, and slightly increases in week 4. The pattern for the effect of combination nicotine replacement therapy (vs. the nicotine patch only) was similar although the initial increase in the magnitude was not as pronounced but the mediation effect was statistically significant throughout the four weeks post-quit. Examination of the two time-varying effects that make up the mediation effect are also informative. Examination of the bottom row of Figure \ref{fig:RealdataAnalysis2} shows that varenicline (vs. the nicotine patch only) has a negative effect that becomes stronger over the course of the first week. This effect then begins to diminish during the following three weeks. In contrast, the strong negative effect of combination nicotine replacement therapy (vs. the nicotine patch only) on negative affect is apparent at the beginning of week one but then, similar to the varenicline group, diminishes during the following three weeks. Examination of the time-varying effect of negative affect on cessation fatigue reveals that there is a strong positive relationship (i.e., more negative affect results in more cessation fatigue) initially during the first week that diminishes over the following three weeks.

Additionally, the mediation effects for both interventions (compared to the nicotine patch only group) are time-varying. Compared to using nicotine patch only, the effects of varenicline and combination nicotine replacement therapy on cessation fatigue, as mediated by negative affect, are not only negative, but also, time-varying for the four weeks post quit day. Both mediation effects have a narrow confidence band and thus, we can rule out a constant mediation effect over time because we would not be able to fit a flat line over time within the confidence interval.

\section{Discussion}\label{section:discussion}
In this paper, we have described a model for assessing mediation in the context of intensive longitudinal data in which both the mediator and outcome variables are time-varying. This model allows for estimation of time-varying mediation effects. Intensive longitudinal data often arise from the collection of EMA data but may also arise from the collection of data from mobile devices, such as wrist-worn or hip-worn accelerometers. The temporal density of these data allow for more nuanced research questions that cannot be addressed by, for example, averaging over the EMA data and/or assuming that the mediated effect does not vary as a function of time. By allowing mediated effects to vary as a function of time, research questions such as the timing of important mediation effects can be assessed. Thus, our approach may prove useful to other researchers who wish to conduct mediation analysis in the context of intensive longitudinal data. 

The simulation study showed that the proposed bootstrap pointwise confidence intervals contained the true time-varying mediation effect and that the estimated time-varying mediation effect was close to the true time-varying mediation effect. We then applied our approach to examine the mediation effect of three smoking cessation treatments (i.e., varenicline, combination nicotine replacement therapy, and nicotine patch only) on cessation fatigue via negative affect. The results indicated that the mediated effect 1) did indeed vary as a function of time, 2) was statistically different from zero throughout the four weeks post-quit day, and 3) that the effect was strongest in the first week post-quit for the varenicline group (vs. nicotine patch only). That is, the varenicline group experienced decreased negative affect during the first week, leading to decreased cessation fatigue. Interestingly, the effect was also strongest in the first week for the patch only group and the effect was immediate whereas the varenicline effect improved over the first half of the first week. The mediated effect for both treatments, compared to the patch alone, appeared to dissipate over the course of the first four weeks of the quit attempt. This information may lead to modifications and/or adaptations of the intervention to, for example, implement a behavioral component to address negative affect, with a specific focus on reducing negative affect in the first week of the quit attempt. This information would not have been evident had we assumed that the mediation effect was invariant across the four week post-quit period.

There are several limitations of the current approach. First, the proposed method only constructs a point-wise confidence interval. For inference at a fixed time point, a point-wise confidence interval is useful. However, a simultaneous confidence band is needed to make inferences over the entire time span. Thus, an obvious future direction is developing and estimating a simultaneous confidence band. Second, although the current approach does not require observations from all participants at all time points \citep{fan2000two}, the algorithm will not work if  $\mbox{rank}(\mathbf{X}^*_j) <3$ (or $<d$ in the general case). In such cases, one can implement the four methods discussed in \cite{fan2000two} (see their Remark 1).

A third limitation is that as with all mediation analyses, we need to make strong assumptions regarding no unmeasured confounding, temporal order, and additivity as mentioned previously. In our particular application, individuals were randomly assigned to the smoking cessation treatments; however, they are not randomly assigned to the mediator and therefore, there may be confounders of the mediator and the outcome. In addition, due to the intensive longitudinal nature of the study, we cannot rule out the possibility of time-varying confounding. Of particular concern is the possibility of time-varying confounders of the mediator and outcome that have themselves been affected by the smoking cessation treatments. In future work, we will propose sensitivity analyses to address potential violations of these assumptions. Additional future work will also address binary outcomes, for example, daily smoking, and count outcomes, for example, daily number of cigarettes smoked.


In conclusion, we have presented a model for estimating time-varying mediation effects which builds on previous work \citep{lindquist2012functional, fan2000two, vanderweele2017mediation} to allow a time-varying outcome as well as a time-varying mediator. We also presented a method for obtaining point-wise confidence intervals for the product of two time-varying coefficient functions (i.e., a time-varying mediation effect), evaluated its feasibility in a small simulation study, and applied the method to evaluate the time-varying mediation effects of three pharmacotherapy smoking cessation interventions. We have implemented the estimation and bootstrap procedure in a user-friendly R package, \emph{tvmediation}, available on GitHub. Although we cannot share the actual data, the R package contains data simulated to mimic the real data along with tutorials on how to use the functions to fit the model described above. We believe that this approach will be useful to those collecting frequent data using mobile devices for self-reported EMA and who wish to examine mediation effects.

\nocite{*}
\bibliographystyle{rss}
\bibliography{mylibrary}

\end{document}